\documentclass[aps,twocolumn,groupedaddress,showpacs]{revtex4}
\usepackage{graphicx}
\usepackage{amsfonts}
\usepackage{amssymb}
\usepackage{amsmath}
\usepackage{wrapfig}

\begin{document}

\title
{Weak correlation effects in the Ising model on triangular-tiled hyperbolic lattices}

\author{Andrej~Gendiar$^1$, Roman Krcmar$^2$, Sabine Andergassen$^3$, 
Michal Dani\v{s}ka$^1$, and Tomotoshi Nishino$^4$}
\affiliation{
$^1$Institute of Physics, Slovak Academy of Sciences, SK-845~11, Bratislava, Slovakia\\
$^2$Physikalisch-Technische Bundesanstalt, D-38116 Braunschweig, Germany\\
$^3$Faculty of Physics, University of Vienna, Boltmanngasse 5, A-1090 Vienna, Austria\\
$^4$Department of Physics, Graduate School of Science, Kobe University,
Kobe 657-8501, Japan}

\begin{abstract}
The Ising model is studied on a series of hyperbolic two-dimensional lattices 
which are formed by tessellation of triangles on negatively curved surfaces. 
In order to treat the hyperbolic lattices, we propose a generalization of the 
corner transfer matrix renormalization group method using a recursive 
construction of asymmetric transfer matrices. Studying the phase transition, 
the mean-field universality is captured by means of a precise analysis of
thermodynamic functions. The correlation functions and the density matrix 
spectra always decay exponentially even at the transition point, whereas power
law behavior characterizes criticality on the Euclidean flat geometry.
We confirm the absence of a finite correlation length in the limit of infinite
negative Gaussian curvature.
\end{abstract}

\pacs{05.50.+q, 05.70.Jk, 64.60.F-, 75.10.Hk}

\maketitle

\section{Introduction}

An increasing interest in the thermodynamic behavior of various physical 
models on non-Euclidean (curved) surfaces has been persisting for about
two decades, due to recent experimental fabrication of soft materials
with conical geometry~\cite{cn} and magnetic nano\-structures which exhibit
negatively curved geometries~\cite{experiment1,experiment2,experiment3}.
Curved geometries are also relevant in the theory of quantum
gravity~\cite{q-gravity1,q-gravity2}. In this context, several
statistical models have been investigated on simple negatively curved
geometries, such as the Ising model~\cite{Shima,hctmrg-Ising-5-4,Sakaniwa},
the $q$-state clock models~\cite{hctmrg-clock-5-4,Baek-clock}, and the
XY-model~\cite{XY-model}.

\begin{figure}[tb]
\centerline{\includegraphics[width=0.5\textwidth,clip]{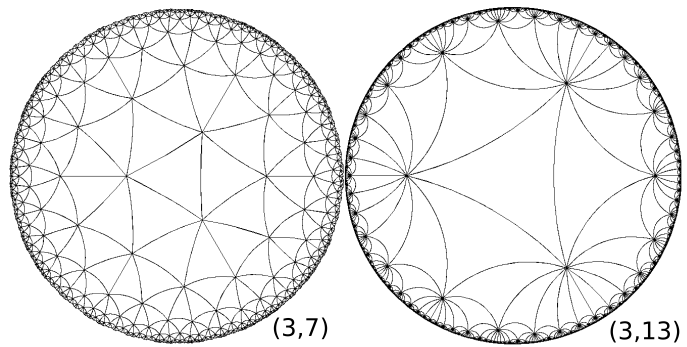}}
\caption{Poincar\'e disk representation of the hyperbolic  lattices
created by triangular tessellation $p=3$ with the coordination
numbers $q=7$ (left) and $q=13$ (right).}
\label{fig1}
\end{figure}

A typical example of the negatively curved geometry is represented by the
two-dimensional discretized hyperbolic surface (lattice) which is characterized
by a constant negative Gaussian curvature. Among the varieties of lattice surfaces,
we choose, for simplicity, a group of regular lattices that are constructed as
tiling of congruent polygons of the $p$-th order with the coordination number
$q$. On the hyperbolic $( p, q )$ lattices, the relation $(p-2)(q-2)>4$ is
satisfied, in contrast to the relation $(p-2)(q-2)=4$ on the Euclidean flat
geometry. Figure \ref{fig1} shows two examples, the $( 3, 7 )$ and $( 3, 13 )$
lattices where the whole lattice is mapped onto the Poincar\'e disk~\cite{Poincare}.

In general, the number of the lattice sites within a certain area increases
exponentially with its diameter on such hyperbolic lattices. This exponential
increase limits efficiency of numerical studies of statistical models, such as
the Ising model on the $( p, q )$ lattice. In particular, applications of Monte 
Carlo simulation face difficulties in the scaling analysis around the phase 
transition. Also transfer matrix diagonalization can not easily be applied due
to the non-triviality in the construction of the row-to-row transfer matrices.

Despite these difficulties, one can evaluate the partition function by means of
Baxter's corner transfer matrix formalism~\cite{Baxter} even for the hyperbolic
$( p, q )$ lattices. In this article, we use a flexible numerical implementation
of Baxter's method, so-called the Corner Transfer Matrix Renormalization Group
(CTMRG) algorithm, which has been used as a tool in the computation of the
partition function for (flat) two- and three-dimensional classical spin 
systems~\cite{ctmrg-tn,ctmrg-snapshot,tpva-Potts}. In our previous
reports~\cite{hctmrg-Ising-5-4,hctmrg-Ising-p-4,hctmrg-clock-5-4,hctmrg-J1J2}
we considered the hyperbolic $( p, q )$ lattices, typically for the case with
$q = 4$, where the whole lattice can be divided into four quadrants, the `corners'.
For the Ising model on the $( p, 4 )$ lattices, the mean-field universality was
found~\cite{Shima,hctmrg-Ising-5-4}.

The hyperbolic $( p, q )$ lattice with an {\it arbitrary} coordination number $q$
other than four has not been addressed by use of the CTMRG method yet. For this
case the numerical renormalization procedure of the corner transfer matrices
requires a technical extension upon the established numerical procedure for the
$( p, 4 )$ lattices. In this article we introduce a new procedure which is valid
for general values of $q$ and find the thermodynamic properties of the Ising model
on a wider class of the $( p, q )$ lattices. In particular, the triangular
tessellation ($p = 3$) and the coordination number $q \geq 6$ are investigated
as representative examples.

This article is organized as follows. In Sec.~II we define the Ising model on
the $( p, q )$ lattices. In Sec.~III the recurrent renormalization algorithm
of the CTMRG method is introduced. The application of CTMRG to the $(3, q)$ 
lattices is explained starting from $q = 6$ and increasing $q$. Numerical
results on the spontaneous magnetization and energy are presented in Sec.~IV, 
with a detailed analysis of the $q$-dependence of the phase transition 
temperature and the corresponding scaling exponents. In Sec.~V the quantum
entropy and the scaling behavior of the correlation functions are observed.
We also analyze the effects of the Gaussian curvature on the correlation length.
We summarize the result in the last section.

%%%%%%%%%%%%%%%%%%%%%%%%%%%%%%%%%%%%%%%%%%%%%%%%%%%
\section{The lattice model}
%%%%%%%%%%%%%%%%%%%%%%%%%%%%%%%%%%%%%%%%%%%%%%%%%%%

Consider the Ising model with the Hamiltonian
\begin{equation}
 {\cal H}(\sigma) = -J\sum\limits_{\{i,j\}}\sigma_i\sigma_j
                    -h\sum\limits_{\{i\}}\sigma_i
\end{equation}
defined on the hyperbolic $( p, q )$ lattices. We here use the standard notation
$(p,q)$ where the first integer $p$ corresponds to the regular polygons with
$p$ sides (or vertices) and where the second one $q$ stands for the coordination
number which is the number of polygons meeting in each vertex. Throughout this
article we focus on the triangular tiling on the $( 3, q )$ lattices only. The
Ising spin variables $\sigma_i^{~} = \uparrow{\rm or } \downarrow$ are located on
the vertices. The first term in ${\cal H}( \sigma )$ represents the ferromagnetic 
coupling ($J>0$) between the nearest-neighboring Ising spins $\sigma_i^{~}$ and
$\sigma_j^{~}$, and the second represents the effect of the external magnetic
field $h$. Then, the partition function
\begin{equation}
 {\cal Z} = \sum\limits_{\{\sigma\}}
            \exp\left[-\frac{\cal H(\sigma)}{k_{\rm B}T}\right]
\end{equation}
is given by the sum of the Boltzmann weights over all spin configurations
which are denoted by $\{\sigma\}$. Here, $k_{\rm B}$ and $T$, respectively,
are the Boltzmann constant and the temperature. 

On any ($p,q$) lattice, the Boltzmann weight of the whole system can be
represented as the product of the local Boltzmann weights attributed to 
the particular $p$-gons. In this study we define the local Boltzmann weights
which are consistent with the triangular tessellation ($p = 3$). For a reason
we explain in the following, each local Boltzmann weight ${\cal W}_{\rm B}$
is constructed by a pair of adjacent triangles $\sigma_a\sigma_b\sigma_d$ and
$\sigma_b\sigma_c\sigma_d$ as shown in Fig.~2. The local Boltzmann 
weight ${\cal W}_{\rm B}$ for this pair of the triangles is then given by
\begin{eqnarray}
\nonumber
 {\cal W}_{\rm B}(\sigma_a\sigma_b\sigma_c\sigma_d)=\exp
  \bigg[\frac{J}{2k_{\rm B}T} (\sigma_a\sigma_b+\sigma_b\sigma_c
    +\sigma_c\sigma_d\\
    +\sigma_d\sigma_a+2\sigma_b\sigma_d) + 
    \frac{h}{qk_{\rm B}T} (\sigma_a+2\sigma_b+\sigma_c+2\sigma_d)
  \bigg]\ .
\label{bw}
\end{eqnarray}
The factor $2$ of $2\sigma_b\sigma_d$ arises from the fact that $\sigma_b$ and
$\sigma_c$ are shared by two triangles, under the tessellation of "bi-triangular" 
Boltzmann weights. Also the factor $2$ appears in $2\sigma_b$ and $2\sigma_d$ 
since the effect of external magnetic field $h$ should be counted for both upper
and lower triangles. Under these factorizations, we proceed the calculation by 
the CTMRG method~\cite{ctmrg-tn,hctmrg-Ising-5-4}.

The standard numerical formalism based on the diagonalization of the row-to-row
transfer matrix is not easily applied under hyperbolic geometries. It has been
shown that the CTMRG method works as an alternative~\cite{hctmrg-Ising-p-4}
when the $( p, 4 )$ lattice is considered under the condition $p \ge 4$.
Recall that the  $( p, 4 )$ lattice can be divided into four equivalent
quadrants by two perpendicular geodesics, and it is easily understood that
each quadrant corresponds to the corner transfer matrix~\cite{hctmrg-Ising-5-4,
hctmrg-Ising-p-4}. Such division of the whole system is not generally admissible
for the $( 3, q )$ lattices, which is under our interest, in particular, when
$q$ is odd; we tackle this case in the following.

%%%%%%%%%%%%%%%%%%%%%%%%%%%%%%%%%%%%%%%%%%%%%%%%%%%
\section{Recurrent RG scheme}
%%%%%%%%%%%%%%%%%%%%%%%%%%%%%%%%%%%%%%%%%%%%%%%%%%%

\begin{figure}[tb]
\centerline{\includegraphics[width=0.5\textwidth,clip]{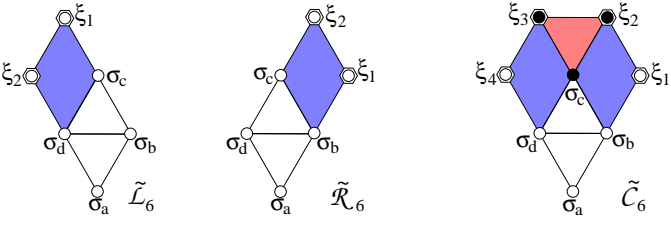}}
\caption{(Color online) Graphical representation of the extension process of
the left transfer matrix ${\tilde{\cal L}}^{~}_6$, the right transfer matrix
${\tilde{\cal R}}^{~}_6$, and the corner transfer matrix ${\tilde{\cal C}}^{~}_6$
on the $( 3, 6 )$ lattice which are defined by Eqs.~\eqref{tl6}--\eqref{c6}.
The filled symbols correspond to the variables which have to be summed up. The
two-state and multi-state variables, respectively, are denoted by $\sigma$ and
$\xi$.}
\label{fig2}
\end{figure}

\begin{figure}[tb]
\centerline{\includegraphics[width=0.5\textwidth,clip]{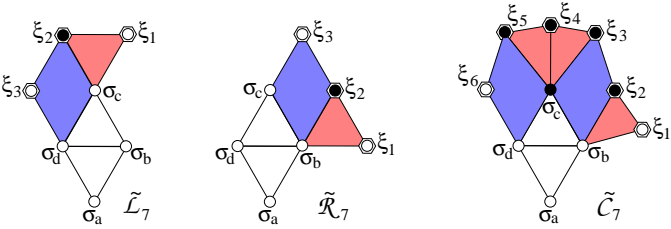}}
\caption{(Color online) The expansion process of 
${\tilde{\cal L}}^{~}_7$, ${\tilde{\cal R}}^{~}_7$, and ${\tilde{\cal C}}^{~}_7$
on the ($3,7$) lattice.}
\label{fig3}
\end{figure}

\begin{figure}[tb]
\centerline{\includegraphics[width=0.5\textwidth,clip]{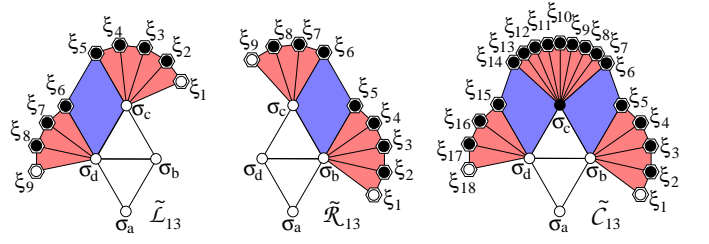}}
\caption{(Color online) The expansion process of ${\tilde{\cal L}}^{~}_{13}$,
${\tilde{\cal R}}^{~}_{13}$, and ${\tilde{\cal C}}^{~}_{13}$ for the ($3,13$)
geometry.}
\label{fig4}
\end{figure}

Let us consider the generalization of the CTMRG method to the $( 3, q )$ lattice 
with $q \ge 6$. The Boltzmann weight of the whole $( 3, q )$ lattice can be 
represented by the product of the $q$ identical corner transfer matrices
${\cal C}_q^{~}$ surrounding the central spin $\sigma$. In this picture we can
express the partition function in the product form
\begin{equation}
\label{pf}
 {\cal Z}_{(p,q)}=\sum\limits_{\sigma}\sum\limits_{\xi_1,\xi_2,\dots,\xi_q}
 \prod\limits_{j=1}^{q} {\cal C}_q(\sigma\xi_j\xi_{j+1}), 
\end{equation}
where $\xi_{j}$ and $\xi_{j+1}$, which appear as the parameters of the corner
transfer matrix ${\cal C}_q(\sigma\xi_j\xi_{j+1})$, are the block spin variables
corresponding to chains of the spins from the central spin $\sigma$ towards
the system boundary. We have assumed the cyclic order around $\sigma$, and
thus $\xi_{q+1}\equiv\xi_{1}$ is satisfied. Throughout this paper we use
the counterclockwise index ordering for the spin variables included in corner 
transfer matrices ${\cal C}_q(\sigma\xi_j\xi_{j+1})$, starting from any one of the 
two-state variables from $\sigma_a$ to $\sigma_d$. Also the renormalized spin 
variables $\xi_i$ are aligned in the same ordering, as shown on the red triangles 
in Figs.~\ref{fig2}, \ref{fig3}, and \ref{fig4}.

Let us explain the recursive construction of the corner transfer 
matrix ${\cal C}_q(\sigma\xi_j\xi_{j+1})$ with respect to the successive area 
expansion of the whole system~\cite{hctmrg-Ising-5-4}. For a tutorial purpose, 
we start from the $( 3, 6 )$ lattice where the system is on the flat surface, and 
treat the cases $q > 6$ afterward. 

In contrast to the original CTMRG formulation~\cite{ctmrg-tn}, it is important to
introduce two different kinds of `half-row transfer matrices' 
${\cal L}_q$ and ${\cal R}_q$ which are used for the area expansion of the 
corner transfer matrix.  
%(See Fig.~2.) 
On the $( 3, 6 )$ lattice, the area expansions of the transfer 
matrices  ${\cal L}_6$ and ${\cal R}_6$ are performed as
\begin{eqnarray}
\label{tl6}
 {\tilde{\cal L}}_6(\sigma_d\sigma_a\sigma_b\sigma_c\xi_1\xi_2)
 ={\cal W}_{\rm B}(\sigma_a\sigma_b\sigma_c\sigma_d)
  {\cal L}_6(\sigma_d\sigma_c\xi_1\xi_2),\\
\label{tr6}
 {\tilde{\cal R}}_6(\sigma_c\sigma_d\sigma_a\sigma_b\xi_1\xi_2)
 ={\cal W}_{\rm B}(\sigma_a\sigma_b\sigma_c\sigma_d)
  {\cal R}_6(\sigma_c\sigma_b\xi_1\xi_2)
\end{eqnarray}
where the position of each spin variable is graphically depicted in Fig.~\ref{fig2}. 
Similarly, the corner transfer matrix ${\tilde{\cal C}}_6$ are expanded as
\begin{eqnarray}
\nonumber
{\tilde{\cal C}}_6(\sigma_d\sigma_a\sigma_b\xi_1\xi_4)
 =\hspace{-0.2cm}\sum\limits_{\sigma_c,\xi_2,\xi_3}\hspace{-0.2cm}
       {\cal W}_{\rm B}(\sigma_a\sigma_b\sigma_c\sigma_d)
       {\cal L}_6(\sigma_d\sigma_c\xi_3\xi_4)\\
 \times{\cal C}_6(\sigma_c        \xi_2\xi_3)
       {\cal R}_6(\sigma_c\sigma_b\xi_1\xi_2) \, .
\label{c6}
\end{eqnarray}
The recursive expansion procedure in CTMRG can be 
initiated by setting
${\cal L}_6(\sigma_a\sigma_b\sigma_c\sigma_d)$
$={\cal R}_6(\sigma_d\sigma_a\sigma_b\sigma_c)$
$={\cal W}_{\rm B}(\sigma_a\sigma_b\sigma_c\sigma_d)$ and
${\cal C}_6(\sigma_a\sigma_b\sigma_d)$
$=\sum_{\sigma_c} {\cal W}_{\rm B}(\sigma_a\sigma_b\sigma_c\sigma_d)$ where
the multi-spin variables $\xi$ are identical with the Ising ones $\sigma$ at
the beginning. In the following, we do not write spin variables explicitly
for book keeping.

We now generalize the above-mentioned expansion process for the $( 3, q )$ lattices,
when $q \geq 7$, where the hyperbolic surface geometry is realized. Drawing the
lattice, such as shown in Fig.~1, and analyzing the inner structure of the corners, 
one can derive a set of recursive relations. The $q$-dependent corner transfer matrix 
\begin{equation}
\label{cq}
 {\tilde{\cal C}}^{~}_q=\sum\limits_{\sigma_c,\xi{\rm 's}}
 {\cal W}_{\rm B}{\cal L}^{~}_q{\cal C}^{~}_q{\cal R}^{~}_q
\end{equation}
is a slight modification of Eq.~\eqref{c6}. The relation is graphically shown 
in Figs.~\ref{fig3} and \ref{fig4} for the two representative cases.
Similarly, for the `half-row transfer matrices', we obtain 
\begin{eqnarray}
\label{lq}
 {\tilde{\cal L}}_q=\sum\limits_{\xi{\rm 's}}{\cal W}_{\rm B}
 {\cal C}^{n_q}_q{\cal L}^{~}_q{\cal C}^{n_q}_q,\\
 {\tilde{\cal R}}_q=\sum\limits_{\xi{\rm 's}}{\cal W}_{\rm B}
 {\cal C}^{n_q}_q{\cal R}^{~}_q{\cal C}^{n_q}_q
\label{rq}
\end{eqnarray}
where $n_q$ is the multiplicity of ${\cal C}^{~}_q$ given by
\begin{equation}
\label{nq}
 n_q=\left\lfloor \frac{q-6}{2} \right\rfloor \equiv
\max\left\{n\in \mathbb{Z} \ \vert \ n\leq \frac{q-6}{2}\right\}.
\end{equation}
In contrast to Eqs.~5 and 6, the corner transfer matrices
appear in the expansion process of ${\cal L}_q$ and ${\cal R}_q$ 
when $q \ge 7$. The extended transfer matrices ${\tilde{\cal L}}^{~}_q$,
${\tilde{\cal R}}^{~}_q$, and ${\tilde{\cal C}}^{~}_q$ re-enter
the right hand sides of Eqs.~\eqref{cq}-\eqref{rq}.

The expansion process successively increases the system size by expanding the
matrix dimensions of  ${\tilde{\cal L}}^{~}_q$, ${\tilde{\cal R}}^{~}_q$,
and ${\tilde{\cal C}}^{~}_q$. To prevent the exponential grow of computational
effort, we introduce the density-matrix renormalization scheme~\cite{ctmrg-tn,
hctmrg-Ising-5-4}. Let us express the block-spin transformation by the matrix
$U_{\rm RG}$. The transfer matrices are `compressed' by the RG transformation
\begin{eqnarray}
\nonumber
\left(U^\dagger_{\rm RG}{\tilde{\cal L}}_q U^{~}_{\rm RG}\right)/
{\vert\vert U^\dagger_{\rm RG}{\tilde{\cal L}}_q U^{~}_{\rm RG}\vert\vert}_2
&\to&{\cal L}_q,\\
\label{rg}
\left(U^\dagger_{\rm RG}{\tilde{\cal R}}_q U^{~}_{\rm RG}\right)/
{\vert\vert U^\dagger_{\rm RG}{\tilde{\cal R}}_q U^{~}_{\rm RG}\vert\vert}_2
&\to&{\cal R}_q,\\
\nonumber
\left(U^\dagger_{\rm RG}{\tilde{\cal C}}_q U^{~}_{\rm RG}\right)/
{\vert\vert U^\dagger_{\rm RG}{\tilde{\cal C}}_q U^{~}_{\rm RG}\vert\vert}_2
&\to&{\cal C}_q \, .
\end{eqnarray}
We introduced the normalization factor ${\vert\vert\cdot\vert\vert}_2$ in
order to avoid the numerical over-flow in the expression of the partition
function. 

The central issue concerns the definition of the RG transformation. In
the density-matrix renormalization scheme, $U_{\rm RG}$ is created by
diagonalization of the reduced density matrix $\rho$ which may be
represented in a non-Hermitian (asymmetric) form
\begin{equation}
\rho={\rm Tr}_{\rm env}\vert\psi\rangle\langle\phi\vert \, .
\label{rdm1}
\end{equation}
The trace is taken over the spin variables belonging to the environment
as proposed by DMRG~\cite{ctmrg-tn,white}. The states $\vert\psi\rangle$
and $\vert\phi\rangle$ correspond to two parts of the whole lattice.
The Boltzmann weight for these two parts can be calculated as the
product of the corner transfer matrices
\begin{eqnarray}
\psi(\sigma\xi_\alpha\xi_\beta)&=&\hspace{-0.5cm}
  \sum\limits_{\xi_1\xi_2,\dots,\xi_k}\hspace{-0.3cm}
        {\cal C}(\sigma\xi_\alpha\xi_1)
        {\cal C}(\sigma\xi_1\xi_2)
  \cdots{\cal C}(\sigma\xi_k\xi_\beta),  \\
\phi(\mu\xi_\gamma\xi_\delta)&=&\hspace{-0.5cm}
  \sum\limits_{\xi_1\xi_2,\dots,\xi_\ell}\hspace{-0.3cm}
        {\cal C}(\mu\xi_\gamma\xi_1)
        {\cal C}(\mu\xi_1\xi_2)
  \cdots{\cal C}(\mu\xi_\ell\xi_\delta)\, ,
\end{eqnarray}
where we  introduced the condition $k+\ell+2=q$. The most optimal choice
is to consider $k=n_{q+5}$ and $\ell=n_{q+4}$ with $n_q$ given by
Eq.~\eqref{nq}. We have used letter
$\mu$ for the 2-state variable of $\phi$ just for distinction from $\sigma$ 
of $\psi$, and this choice is convenient when we construct the reduced 
density matrix. The normalized partition function can be written as
${\cal Z}_{(3,q)}=\langle\psi\vert\phi\rangle$. As an example, we obtain
$k=3$ and $\ell=2$ for the $(3,7)$ lattice with the corresponding
Boltzmann weights
\begin{eqnarray}
\nonumber
\psi(\sigma\xi_\alpha\xi_\beta)&=&\sum_{\xi_1\xi_2\xi_3} {\cal C}(\sigma
\xi_\alpha\xi_1){\cal C}(\sigma\xi_1\xi_2){\cal C}(\sigma\xi_2\xi_3){\cal C}
(\sigma\xi_3\xi_\beta)\, ,\\
\phi(\mu\xi_\gamma\xi_\delta)&=&\sum_{\xi_1\xi_2}{\cal C}
(\mu\xi_\gamma\xi_1){\cal C}(\mu\xi_1\xi_2){\cal C}(\mu\xi_2\xi_\delta)\, .
\end{eqnarray}

Notice that if $q$ is even, $k=\ell=\frac{q}{2}-1$ resulting in
$\vert\psi\rangle\equiv\vert\phi\rangle$. For this choice the reduced
density matrix $\rho$ is always Hermitian (symmetric). However,
for any odd $q$, $\rho$ becomes non-Hermitian (asymmetric). This
may lead to severe numerical instabilities. In order to avoid them,
we symmetrize the reduced density matrix. We, therefore, consider an
equally weighted reduced density matrix
\begin{eqnarray}
\nonumber
\rho(\sigma\xi_\alpha\vert\mu\xi_\beta)=\frac{1}{2}&\sum\limits_{\xi_\gamma}
   & \psi^\dagger(\sigma\xi_\alpha\xi_\gamma)\phi(\mu\xi_\beta\xi_\gamma)\\
   &+&\phi^\dagger(\sigma\xi_\alpha\xi_\gamma)\psi(\mu\xi_\beta\xi_\gamma)\, .
\label{rdm2}
\end{eqnarray}
Having tested both formulations of the reduced density matrix,
Eqs.~\eqref{rdm1} as well as \eqref{rdm2}, we encountered numerical
instabilities for the non-Hermitian case only, especially in the
vicinity of the phase transition. Otherwise, both density matrix
formulations yield the identical thermodynamic properties.

%%%%%%%%%%%%%%%%%%%%%%%%%%%%%%%%%%%%%%%%%%%%%%%%%%%
\section{Magnetization and Energy}
%%%%%%%%%%%%%%%%%%%%%%%%%%%%%%%%%%%%%%%%%%%%%%%%%%%

Since the detailed analysis of the phase transitions deep inside
the hyperbolic lattice is of our interest, we concentrate on the
bulk properties of a sufficiently large inner region of the
lattice~\cite{Sakaniwa,hctmrg-Ising-p-4} although the influence
of the system boundary is not negligible at all for the
discussion of the thermodynamic properties of the system. The bulk
spontaneous magnetization is an example where the value can be
calculated by
\begin{equation}
M={\rm Tr} \left( \sigma \rho \right) / {\rm Tr} \,  \rho
\end{equation}
in the CTMRG formulation. Without loss of generality, we set the
coupling constant $J$ and the Boltzmann constant $k_{\rm B}$ to
unity, and all thermodynamic functions are evaluated in the unit
of $k_{\rm B}$.

\begin{figure}[tb]
\centerline{\includegraphics[width=0.5\textwidth,clip]{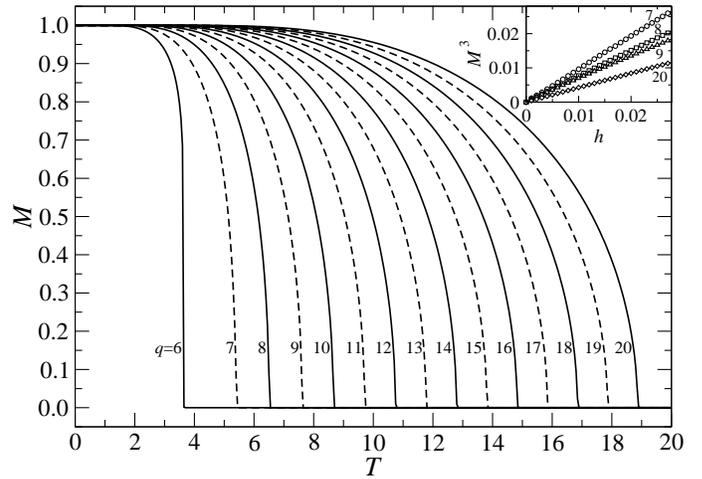}}
\caption{Spontaneous magnetizations $M$ with respect to temperature $T$ for
$6\leq q \leq 20$. The inset shows the linear behavior of the cubic power of
the induced magnetization $M^3_{~}$ with respect the magnetic field $h$
around the transition temperatures $T_{\rm pt}^{(q)}$.}
\label{fig5}
\end{figure}

We now consider one-point functions of the Ising model on the series of
$( 3, p )$ lattices in the thermodynamic limit. First of all, let us check
the validity of our numerical procedure as explained in the previous section.
We perform a test calculation for the flat $( 3, 6 )$ lattice. Keeping only
$m = 20$ states of the multi-spin variables $\xi$~\cite{hctmrg-Ising-5-4,
hctmrg-Ising-p-4,white}, the obtained spontaneous magnetization is shown in
Fig.~\ref{fig5}. The estimated transition temperature $T_{\rm c}=3.641$ is
quite close to the exact value $T_{\rm c}=4/\ln 3\approx3.64096$~\cite{Baxter}.

Now, we focus on the hyperbolic surfaces. In Fig.~\ref{fig5}
we also plot the temperature dependence of the spontaneous magnetization $M$
for the coordination numbers from $7\leq q \leq 20$. The
full and the dashed curves, respectively, distinguish the even and odd values
of $q$. 
% Table~\ref{t1} lists the resulted transition temperatures $T_{\rm pt}^{(q)}$. 
As we show later, the system is always off-critical whenever
$q \geq 7$, even at the transition temperature. We, therefore, use the
notation $T_{\rm pt}^{(q)}$ instead of $T_{\rm c}^{(q)}$ for $q \geq 7$;
we also use $T_{\rm pt}^{(6)}$ for $q = 6$ in order to unify the notation.

% \begin{table}[bt]
% \begin{center}
% \begin{tabular}{|c|r||c|r||c|r|}
% \hline
% ($p,q$)   &  $T_{\rm pt}^{(q)}$\   &
% ($p,q$)   &  $T_{\rm pt}^{(q)}$\ \ &
% ($p,q$)   &  $T_{\rm pt}^{(q)}$\ \ \\ \hline
% $( 3, 6 )$   &  3.641  &  ($3,11$)  &   9.709  &  ($3,16$)  &  14.822 \\ \hline
% ($3,7$)   &  5.407  &  ($3,12$)  &  10.742  &  ($3,17$)  &  15.837 \\ \hline
% ($3,8$)   &  6.511  &  ($3,13$)  &  11.770  &  ($3,18$)  &  16.846 \\ \hline
% $( 3, 6 )$   &  7.608  &  ($3,14$)  &  12.789  &  ($3,19$)  &  17.859 \\ \hline
% ($3,10$)  &  8.633  &  ($3,15$)  &  13.809  &  ($3,20$)  &  18.865 \\ \hline
% \end{tabular}
% \end{center}
% \caption{Transition temperatures of the Ising model on the
% $( 3, q )$ lattices.}
% \label{t1}
% \end{table}

If a small magnetic field $h$ is applied at the transition temperature 
$T_{\rm pt}^{( q \ge 7 )}$, the cubed induced magnetization $M^3_{~}$ is
always linear around $h = 0$. Thus, the model satisfies the scaling relation
$M(h,T_{\rm pt})\propto h^{1/\delta}$ with the scaling exponent $\delta=3$. 
This value is known for the mean-field universality of the Ising model 
and is in full agreement with our previous results for the 
hyperbolic $( p \ge 5, 4 )$ lattices~\cite{hctmrg-Ising-p-4}.

\begin{figure}[tb]
\centerline{\includegraphics[width=0.5\textwidth,clip]{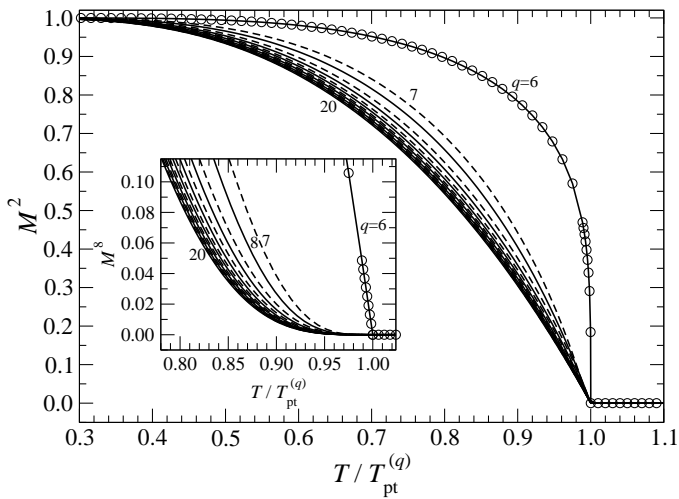}}
\caption{The squared spontaneous magnetization $M^2_{~}$ is linear with 
respect to the normalized temperature near the transition point. 
This corresponds to $\beta=\frac{1}{2}$. 
Inset: the linearity of the $M^8$ is observed only when $q=6$ where $\beta=\frac{1}{8}$.}
\label{fig6}
\end{figure}

\begin{figure}[tb]
\centerline{\includegraphics[width=0.5\textwidth,clip]{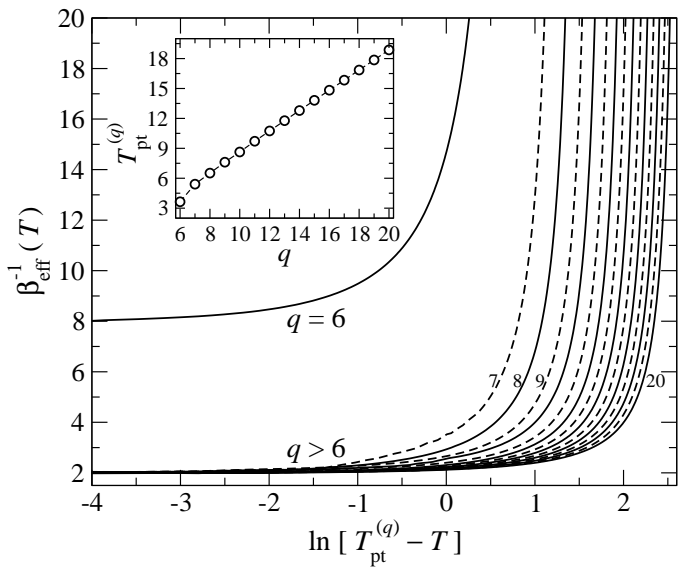}}
\caption{Convergence rate of the effective scaling exponent. Inset:
scaling of the phase transition temperatures $T_{\rm pt}^{(q)}$ versus the
integer $q$.}
\label{fig7}
\end{figure}

In order to observe the scaling relation of the spontaneous magnetization $M$
in a unified manner, we plot the squared magnetization $M^2_{~}$ in Fig.~\ref{fig6}
with respect to the rescaled temperature by $T_{\rm pt}^{(q \ge 7)}$.
Near the point $T = T_{\rm pt}^{(q \ge 7)} $ the
mean-field behavior $M(h=0,T)\propto(T_{\rm pt}^{(q)} - T)^\beta_{~}$ 
with  $\beta=\frac{1}{2}$ is detected. Note that on the $( 3, 6 )$ lattice
the exponent is $\beta=\frac{1}{8}$ as displayed in the inset.
To detect the scaling exponent $\beta$ in a more precise manner,
we calculate the effective exponent
\begin{equation}
\beta_{\rm eff}(T)=\frac{\partial \ln \left[ M \left( h=0,T<T_{\rm pt}^{(q)}
      \right) \right]} {\partial \ln \left[T_{\rm pt}^{(q)} - T\right]} \, 
\label{beff}
\end{equation}
%
%
% From here below, Sabine. Thank you. :-)
% yes!       : )
%
by means of the numerical derivative. The convergence of $\beta_{\rm eff}(T)$
with respect to $T_{\rm pt}^{(q)} - T$ is shown in Fig.~\ref{fig7}.  It is
apparent that the mean-field value $\beta=\frac{1}{2}$ is detected for any
$q \ge 7$, whereas we confirm $\beta=\frac{1}{8}$ on the flat $(3,6)$ lattice
only. The linear increase of the transition temperature $T_{\rm pt}^{(q \ge 7)}$ 
with respect to $q$ is shown in the inset where the linearity appears already 
around $q\gtrsim8$. This agrees with an intuition where the mean-field behavior
becomes dominant for large coordination numbers.

\begin{figure}[tb]
\centerline{\includegraphics[width=0.5\textwidth,clip]{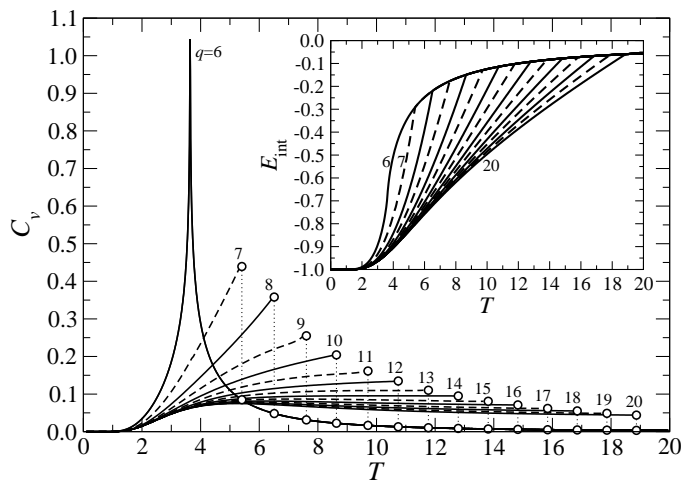}}
\caption{Specific heat as a function of temperature. The open circles connected
by the vertical dotted lines show the discontinuity. Inset: temperature 
dependence of the internal energy.}
\label{fig8}
\end{figure}

Let us analyze the specific heat (or the heat capacity) per bond
\begin{equation}
C_v^{~}=\frac{\partial E_{\rm int}^{~}}{\partial T}
\end{equation}
where $E_{\rm int}^{~}$ is the internal energy per bond, or equivalently,
the correlation function between the two nearest-neighbor spins 
\begin{equation}
E_{\rm int}^{~} =-J\langle\sigma_{i}\sigma_{i+1}\rangle
           =-J\,{\rm Tr}\left(\sigma_{i}\sigma_{i+1}\rho\right)
\end{equation}
with $\sigma_{i}$ and $\sigma_{i+1}$ located at the center of the lattice.
Figure~\ref{fig8} shows the results for $C_v^{~}$ and $E_{\rm int}^{~}$. 
The internal energy $E_{\rm int}^{~}$ is continuous for all the cases we
computed. The presence of the kink in $E_{\rm int}^{~}$ at the
transition temperature for each $q \geq 7$ corresponds to the
discontinuity in $C_v^{~}$~\cite{hctmrg-Ising-p-4,hctmrg-J1J2}.
For these cases the scaling exponent $\alpha$, which appears in the relation
$C_v^{~}(h=0,T)\propto {\vert T_{\rm pt}^{(q)} - T \vert}^{-\alpha}$, is zero.
It is instructive to point out that both $C_v^{~}$ and $E_{\rm int}$ in the
paramagnetic region are almost independent on $q$; the tiny differences are
hardly visible on the scale in the figure.

%%%%%%%%%%%%%%%%%%%%%%%%%%%%%%%%%%%%%%%%%%%%%%%%%%%
\section{Entropy and Correlation}
%%%%%%%%%%%%%%%%%%%%%%%%%%%%%%%%%%%%%%%%%%%%%%%%%%%

\begin{figure}[tb]
\centerline{\includegraphics[width=0.5\textwidth,clip]{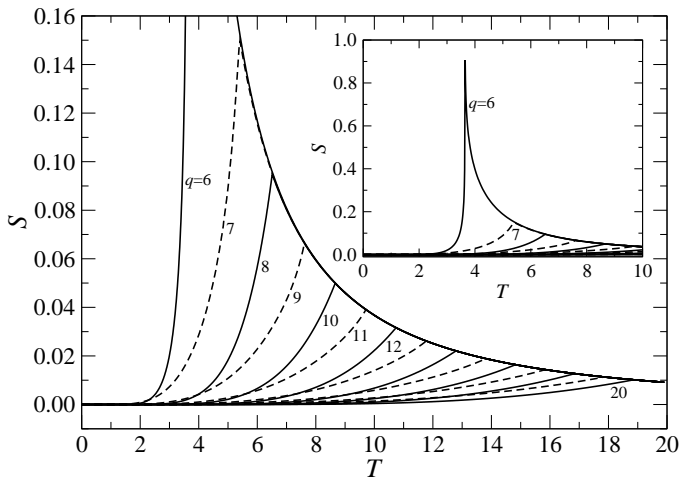}}
\caption{Temperature dependence of the von Neumann entanglement entropy.
The inset displays the dominant behavior of $S$ for the $( 3, 6 )$ lattice.}
\label{fig9}
\end{figure}

Whenever the reduced density matrix $\rho$ is defined, 
the von Neumann (or entanglement) entropy~\cite{rem1} 
\begin{equation}
 S=-{\rm Tr} \left(\rho\log_2\rho\right)
  =-\sum\limits_{i}\omega_i^{~}\log_2\omega_i^{~}
\end{equation}
can be used as a characteristic quantity which is of use for the classification 
of the phase transition. Figure~\ref{fig9} shows the temperature dependence
of $S$ which remains finite for $q \ge 7$ even at the transition temperature.
The entropies in the paramagnetic region are also almost independent on $q$
if $q \geq 7$ as found for $C_v^{~}$ and $E_{\rm int}$.

\begin{figure}[tb]
\centerline{\includegraphics[width=0.5\textwidth,clip]{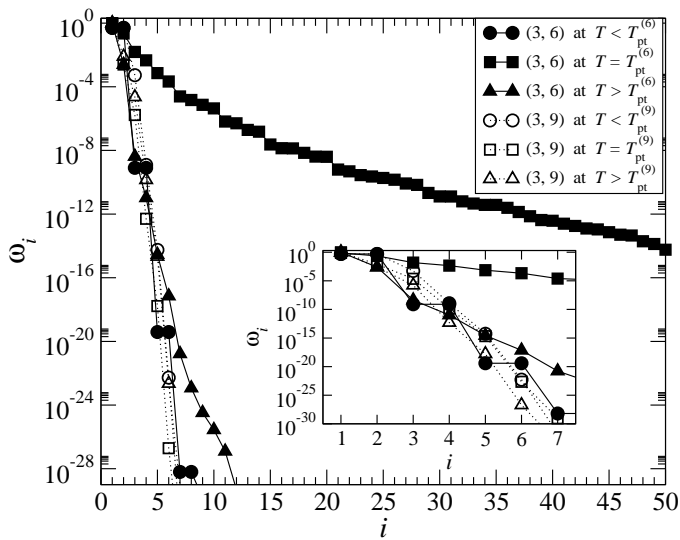}}
\caption{Decay of the density matrix spectra for the $( 3, 6 )$ lattice
(filled symbols) and the $( 3, 9 )$ lattice (open symbols).}
\label{fig10}
\end{figure}

The decay rate of the density matrix eigenvalues $\omega_i^{~}$ is shown in
Fig.~\ref{fig10} on a semilogarithmic scale for both $(3, 6)$ and $(3, 9)$
lattices. We confirm a power-law decay in $\omega_i^{~}$ only at the
transition point of the $( 3, 6 )$ lattice. Note that the
eigenvalues $\omega_i^{~}$ decrease exponentially for $q\geq7$ at the
transition temperature.

The exponential decay of the density matrix spectra is also reflected in the
correlation function
\begin{equation}
G_{i,j}={\rm Tr}\left( \sigma_i \sigma_j \rho \right)
\label{cf}
\end{equation}
between two distant sites $i$ and $j$. We place the spin $\sigma_i$ at the
center of the system and $\sigma_j$ at the system boundary. Therefore, as the
lattice expands its size via the recursive steps in CTMRG, the distance between
these two spins increases progressively.

\begin{figure}[tb]
\centerline{\includegraphics[width=0.5\textwidth,clip]{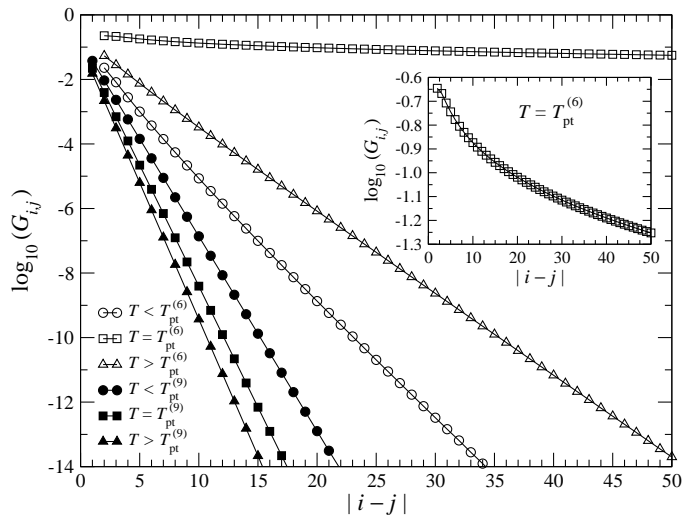}}
\caption{Decay of the correlation functions with respect to the
distance $\vert i-j\vert$. The open and the full symbols, respectively,
correspond to the  $( 3, 6 )$ lattice calculated at $T=3.0$, $3.641$,
and $5.0$, and the $( 3, 9 )$ lattice at $T=6.0$, $7.608$, and $9.0$.}
\label{fig11}
\end{figure}

Figure~\ref{fig11} depicts $\log_{10} \left(G_{i,j}\right)$ as a function of
$\vert i-j\vert$ for the $( 3, 6 )$ lattice (open symbols) and the  $( 3, 9 )$
lattice (full symbols). It is evident that the correlation functions always
decay exponentially on the $( 3, 9 )$ lattice regardless of the temperature.
We remark that an analogous exponential decay of $G_{i,j}$ has been observed
for all $q \ge 7$ (not shown). On the $( 3, 6 )$ lattice, the correlation
function decays as a power law at the transition temperature, as seen in the
inset. 

In the following, we compare the Gaussian curvature associated to the $(3, q)$
lattice with the correlation length at the transition temperature. There
are several ways to define the correlation length $\xi_q$~\cite{Baxter,Frank}.
For example, the decay rate of the correlation function directly provides $\xi_q$. 
This is straightforward, but the region of the distance for the fitting analysis
has to be valued carefully. Another possibility consists in using the largest
eigenvalue $\lambda_0(q)$ and the second largest one  $\lambda_1(q)$ of the
row-to-row transfer matrix where $\xi_q$ is determined from
\begin{equation}
\frac{1}{  \xi_q} = \ln\left[\frac{\lambda_0(q)}{\lambda_1(q)}\right] \, .
\label{corlen}
\end{equation}
The relation can be generalized to the $( 3, q \ge 7 )$ lattices, in analogy
to our previous formulations for  the $( 5, 4 )$ lattice~\cite{hctmrg-corrlen}, 
via the construction of the row-to-row transfer matrix
\begin{equation}
  {\cal T}_q(\xi_1 \sigma_a \xi_2|\xi_1^\prime \sigma_a^\prime \xi_2^\prime)
   ={\cal L}_q(\sigma_a^\prime \xi_1^\prime \xi_1 \sigma_a)
    {\cal L}_q(\sigma_a \xi_2 \xi_2^\prime \sigma_a^\prime) \, .
\label{TrMat}
\end{equation}
Using the notation of the recurrence scheme introduced in the
previous section, we calculate $\xi_q$ by use of Eq.~(24).

\begin{figure}[tb]
\centerline{\includegraphics[width=0.5\textwidth,clip]{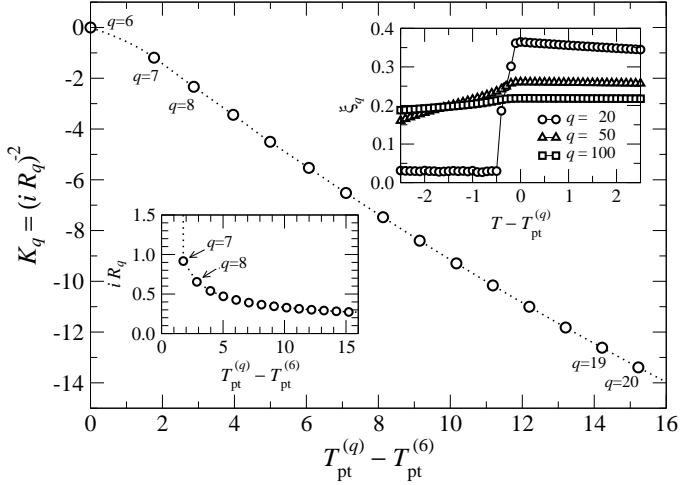}}
\caption{Gaussian curvature $K_q$ with respect to the shifted phase transition
temperatures for $6\leq q\leq 20$. The inset on the left shows the related
radius of the curvature $iR_q$ via Eq.~\eqref{Kq-Rq} while that on the right
shows the correlation length in the vicinity of the phase transition.}
\label{fig12}
\end{figure}

The Gaussian curvature $K_q$ that corresponds to $( 3, q )$ lattice is given
by~\cite{Mosseri}
\begin{equation}
 K_q = \frac{1}{{(iR_q)}^2}=-4\,{\rm arccosh}
  \left[
    \frac{1}{2\sin\left(\frac{\pi}{q}\right)}
  \right]
\label{Kq-Rq}
\end{equation}
where $R_q$ is the curvature radius of the hyperbolic surface.
Recall that $K_q$ must be zero on the Euclidean flat space ($q=6$).
Figure~\ref{fig12} shows the relation between $K_q$ and  the 
shifted transition temperature $T_{\rm pt}^{(q)}-T_{\rm pt}^{(6)}$.
The lower-left inset shows complementary information about $R_q$.
The correlation function $\xi_q$ calculated around the phase transition
for three different $q$'s is plotted in the upper-right inset. Notice
that $\xi_q$ reaches its maximum at the phase transition which is not
well visible as $q$ increases.

\begin{figure}[tb]
\centerline{\includegraphics[width=0.5\textwidth,clip]{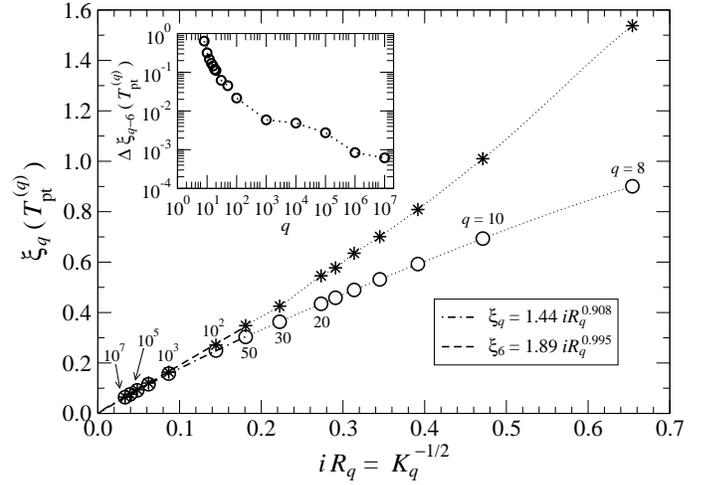}}
\caption{Asymptotic scaling of the correlation length $\xi_q$ at the transition
temperatures $T_{\rm pt}^{(q)}$ with respect to $R_q$. The thin dotted lines
serve as a guide to the eye. The inset shows the difference $\Delta\xi_{6-q}
(T_{\rm pt}^{(q)})$ in Eq.~\eqref{deltaq} with respect to $q$ on a double 
logarithmic scale.}
\label{fig13}
\end{figure}

Figure~\ref{fig13} shows the dependence of the correlation length $\xi_q(T)$ at
the transition temperature with respect to the curvature radius $R_q$. In order
to collect these data, we performed extensive calculations up to 32 digits
numerical precision for the value of $q$ as large as $q = 10\,000\,000$ where
the corresponding Gaussian curvature $K_{10^7}$ is approximately  $900$. Note
that both quantities diverge on the $( 3, 6 )$ lattice, and therefore
$\xi_6(T_{\rm pt}^{(6)})$ and $R_6$ are not shown. Let us focus on the limit
$R_q\to0$ which corresponds to $q \rightarrow \infty$. Evidently, the correlation
length $\xi_q$ decreases to zero as $q$ tends toward infinity (the circles). 
Applying a least-square fit, we obtain the relation $\xi_q=1.44 (iR_q^{~})^{0.908}_{~}$
as shown by the thick dot-dashed curve. If we consider the error in the
calculation of the correlation length, we can conjecture that $\xi_q$ is
proportional to $R_q$.

Recall that the specific heat $C_v^{~}$, the internal energy $E_{\rm int}$,
and the entanglement entropy $S$ turned out to be weakly dependent on the
value of $q$ in the paramagnetic region $T>T_{\rm pt}^{(q)}$ for $q \ge 6$.
Thus, it can be conjectured that the disordered state is not modified by the
presence of the negative curvature. We, therefore, compare $\xi_{q\ge7}$ just
at the transition temperature $T_{\rm pt}^{(q)}$ with the correlation length
$\xi_6$ at the temperatures $T=T_{\rm pt}^{(q)}$. These values are plotted in
Fig.~\ref{fig13} by the asterisks. Since $T_{\rm pt}^{(q)}$ almost linearly
increases with $q$ for large values of $q$, the dotted line goes to the origin of the
graph. The circles and the asterisks in Fig.~\ref{fig13} are of the same order
for all $q$, and this fact supports our conjecture that $R_q$ represents the
only characteristic length of the hyperbolic lattice and that the phase
transition occurs at the temperature where $\xi_q$ is of the same order as
$R_q$. Note that $\xi_6(T_{\rm pt}^{(q)})>\xi_q(T_{\rm pt}^{(q)})$ is always
fulfilled as plotted in the inset of Fig.~\ref{fig13} where we show
the difference
\begin{equation}
\Delta\xi_{6-q}(T_{\rm pt}^{(q)})=
\left[\xi_6(T_{\rm pt}^{(q)})-\xi_q(T_{\rm pt}^{(q)})\right]\, .
\label{deltaq}
\end{equation}
The relation $\xi_6(T_{\rm pt}^{(q)})>\xi_q(T_{\rm pt}^{(q)})$ may be
explained by the effect of the negative curvature that prevents from a kind 
of {\it loop-back} of the correlation effect. Such suppression is also expected
to be present in higher-dimensional hyperbolic lattices and could be analytically
studied by means of the high temperature expansion.

We conjecture the reason why the correlation length remains finite even at the 
phase transition temperature $T_{\rm pt}^{(q)}$ for $q>6$, as follows. First of
all, the hyperbolic plane contains the typical length scale $R_q^{~}$, and it
might prevent scale invariance of the state expected at the criticality. A more
constructive interpretation could be obtained from the observation on the
row-to-row transfer matrix. The calculation of $\xi_q$ by means of Eq.~\eqref{corlen} 
requires diagonalization of the row-to-row transfer matrix ${\cal T}_q(\xi_1\sigma_a\xi_2|
\xi_1^\prime\sigma_a^\prime\xi_2^\prime)$ in Eq.~\eqref{TrMat}. 
The matrix corresponds to an area which connects (transfers) the
row of the neighboring spins $\{\xi_1\sigma_a\xi_2\}$ with the adjacent
ones $\{\xi_1^\prime\sigma_a^\prime\xi_2^\prime\}$. The shape of this area is 
very different from the standard transfer matrix on the Euclidean lattice,
which corresponds to a stripe of constant width. On the hyperbolic surfaces,
however, this distance between the spin rows is not uniform.
The distance is minimal at the center of the transfer matrix, i.e.,
between the two spins $\sigma_a$ and $\sigma_a^\prime$, and it increases 
exponentially with respect to the deviation  from the center to the direction of 
spin rows. Such a geometry~\cite{hctmrg-corrlen} could be imagined from the 
recurrence construction in Eq.~\eqref{lq}. As a consequence, the transfer matrix 
has an effective width, which is of the order of the curvature radius $R_q$. 
The region outside this width contributes as a sort of the boundary spins that
imposes mean-field effect to the bulk part. This situation is analogous to the Bethe 
lattice, being interpreted here as  ($\infty,q$)-lattices.~\cite{hctmrg-Ising-p-4}. 
Thus the Ising  universality could be observed only when the correlation 
length $\xi_q$ is far less than the curvature radius, $\xi_q \llless R_q$. As the
length $\xi_q$ increases toward the transition temperature, we expect a transient 
behavior to the mean-field 
behavior around the point when $\xi_q$ becomes comparable to $R_q$.
We are confirming these conjectures and the details would be reported
in our subsequent work.

\section{Conclusions}

We have presented a detailed analysis of various non-Euclidean lattices forming
surfaces with hyperbolic curvatures. In addition to our previous works on the
($p,4$) lattices, we studied the complementary situation represented by the 
$( 3, q )$ lattices. This task required a reformulation of the existing CTMRG
algorithm. We, therefore, considered the half-row transfer matrices and the
corner transfer matrices including asymmetric (non-Hermitian) cases. For the
lattices with odd $q$'s, we symmetrized the density matrix by the way which
has been accepted by the DMRG community~\cite{white}.

We treated the Ising model on the $( 3, q )$ lattice with coordination number
$6\leq q \leq 10^7$. The phase transition temperatures are determined from
the analysis of the magnetization, internal energy, specific heat, and the
von Neumann entanglement entropy. We have shown that the transition temperature
$T_{\rm pt}^{(q)}$ linearly increases with $q$ for larger values of $q$. The
scaling behavior of the thermodynamic functions, including their related scaling
exponents $\alpha=0$, $\beta=\frac{1}{2}$, and $\delta=3$, obeys the mean-field
universality class. The mean-field nature of the hyperbolic surfaces is
characterized by the exponential decay of the reduced density matrix
eigenvalues and the correlation functions even at the transition temperature. 

We further evaluated the radius of the Gaussian curvature $R_q$ for the generic
($3,q\geq6$) lattice geometry and compare it to the results for the correlation
length extracted from the row-to-row transfer matrix. We found a strongly
suppressed correlation length $\xi_q<1$ at the  transition point for any $q\geq 7$.
We conjecture that $\xi_q$ is proportional to $R_q$ in the large $q$ limit. 

In order to elucidate the origin of the mean-field universality induced by
the hyperbolic geometry, our future studies aim at the treatment of specific
hyperbolic geometries with non-constant Gaussian curvatures in order to
systematically approach the Euclidean (flat) geometry.

\begin{acknowledgments}

A.G. thanks Frank~Verstraete and Vladim\'{\i}r~Bu\v{z}ek for valuable discussions.
This work was supported by the European Union projects meta-QUTE NFP26240120022,
Q-ESSENCE No. 2010-248095, HIP 221889, COQI APVV-0646-10, and VEGA-2/0074/12.
T.~N. acknowledges the support of Grant-in-Aid for Scientific Research.

\end{acknowledgments}


\begin{thebibliography}{99}
\bibitem{rem1} Strictly speaking, the non-Hermitian form of the density matrix
cannot be used to evaluate the entanglement entropy due to the definition of
the density matrix itself. Therefore, we always apply the symmetrized form of
the density matrix for odd $q$'s according to Eq.~\eqref{rdm2}. The entanglement
entropy for odd $q$'s is then considered to be less reliable than for even $q$'s,
and we regard such entropy as complementary information.
\end{thebibliography}
\end{document}